\documentclass[english,reqno]{scrartcl}
\usepackage[T1]{fontenc}
\usepackage[latin9]{inputenc}
\usepackage{geometry}
\usepackage{color}
\usepackage{babel}
\usepackage{units}
\usepackage{amsmath}
\usepackage{amssymb}
\usepackage[unicode=true,pdfusetitle,
 bookmarks=true,bookmarksnumbered=false,bookmarksopen=false,
 breaklinks=false,pdfborder={0 0 0},backref=false,colorlinks=true]
 {hyperref}
\usepackage{breakurl}

\makeatletter
\numberwithin{equation}{section}
\numberwithin{figure}{section}

\usepackage{graphicx}
\usepackage{fancybox}
\usepackage{url}
\usepackage{hyperref}

\makeatother

\begin{document}
\global\long\def\tl#1{\dot{#1}}
\global\long\def\part#1#2{\frac{\partial#1}{\partial#2}}
\global\long\def\adj#1{\bar{#1}}
\global\long\def\p#1{\partial#1}

\title{Adjoints and automatic (algorithmic) differentiation in computational
finance}

\author{Cristian Homescu%
\thanks{Email address: cristian.homescu@gmail.com%
}}

\date{Revised version: May 8, 2011%
\thanks{Original version: May 1, 2011%
}}

\maketitle
Two of the most important areas in computational finance: Greeks and,
respectively, calibration, are based on efficient and accurate computation
of a large number of sensitivities. This paper gives an overview of
adjoint and automatic differentiation (AD), also known as algorithmic
differentiation, techniques to calculate these sensitivities. When
compared to finite difference approximation, this approach can potentially
reduce the computational cost by several orders of magnitude, with
sensitivities accurate up to machine precision. Examples and a literature
survey are also provided.

\pagebreak{}

\tableofcontents{}

\pagebreak{}

\section{Introduction}

Two of the most important areas in computational finance: Greeks and,
respectively, calibration, are based on efficient and accurate computation
of a large number of sensitivities. This paper gives an overview of
adjoint and automatic(algorithmic) differentiation techniques to calculate
those sensitivities. While only recently introduced in the field of
computational finance, it was successfully employed in the last 20
years in other areas, such as computational fluid dynamics, meteorology
and atmospheric sciences, engineering design optimization, etc: \cite{Charpentier_Ghemires_2000,Courtier_Talagrand_1987,Giles_Pierce_2000,Giles_et_al_2006,Homescu_Navon_2003_a,Homescu_Navon_2003_b,LeDimet_et_al_2002,Navon_et_al_1992,Talagrand_Courtier_1987},
to mention but a few.

The computation of the sensitivities is done by performing differentiation
at either continuous level or, respectively, at discrete level. The
{}``continuous'' adjoint approach corresponds to the case where
the adjoint equation is formulated at the differential equation level,
and then discretized. In contrast, the {}``discrete'' adjoint approach
starts with discretized equations, and then formulates the corresponding
discrete adjoint equations. Consequently, the continuous adjoint is
also known under {}``differentiate then discretize'' moniker, while
the discreet adjoint corresponds to {}``discretize then differentiate''
moniker.

The discrete adjoint approach is preferred in many occasions, for
several practical reasons:
\begin{itemize}
\item constructing the discrete adjoint equations is a more straightforward
and systematic process
\item Automatic Differentiation software can be employed to greatly reduce
the development time
\end{itemize}
For this reason we will concentrate in this paper on the discrete
adjoint approach, and for ease of presentation we refer to it as the
adjoint approach. However, since a few papers that are reviewed here
are also based on continuous adjoint, we will make that information
explicit when we refer to those papers, and use the abbreviation ContAdj
for the {}``continuous adjoint'' approach.

Automatic Differentiation (AD), also known as Algorithmic Differentiation,
is a chain-rule-based technique for evaluating the derivatives with
respect to the input variables of functions defined by a high-level
language computer program. AD relies on the fact that all computer
programs, no matter how complicated, use a finite set of elementary
(unary or binary, e.g. sin(·), sqrt(·)) operations as defined by the
programming language. The value or function computed by the program
is simply a composition of these elementary functions. The partial
derivatives of the elementary functions are known, and the overall
derivatives can be computed using the chain rule.

AD has two basic modes of operations, the forward mode and the reverse
mode. In the forward mode the derivatives are propagated throughout
the computation using the chain rule, while the reverse mode computes
the derivatives for all intermediate variables backwards (i.e., in
the reverse order) through the computation. In the literature, AD
forward mode is sometimes referred to as $tangent\, linear$ mode,
while AD reverse mode is denoted as $adjoint$ mode.

The adjoint method is advantageous for calculating the sensitivities
of a small number of outputs with respect to a large number of input
parameters. The for- ward method is advantageous in the opposite case,
when the number of outputs (for which we need sensitivities) is larger
compared to the number of inputs.

When compared to regular methods (such as finite differencing) for
computing sensitivities, AD has 2 main advantages: reduction in computational
time and accuracy. The reduction in computational time is assured
by a theoretical result \cite{Griewank_Walther_2008}that states that
the cost of the reverse mode is smaller than five times the computational
cost of a regular run. The computational cost of the adjoint approach
is independent of the number of inputs for which we want to obtains
the sensitivities with respect to, whereas the cost of thetangent
linear approach increases linearly with the number of inputs. Regarding
accuracy, AD computes the derivatives exactly (up to machine precision)
while finite differences incur truncation errors. The size of the
step $h$ needed for finite difference varies with the current value
of of input parameters, making the problem of choosing $h$, such
that it balances accuracy and stability, a challenging one. AD on
the other hand, is automatic and time need not be spent in choosing
step-size parameters, etc.

AD software packages can also be employed to speed up the development
time. Such tools implement the semantic transformation that systematically
applies the chain rule of differential calculus to source code written
in various programming languages. The generated code can operate in
forward or reverse mode (tangent linear or adjoint model).

\section{General description of the approach}

Let us assume that we have a model dependent on a set of input parameters
which produces an output Y. We also denote by $Z=\left\{ X_{1},X_{2},...,X_{N}\right\} $
the vector of input parameters with respect to which we want to compute
sensitivities.

\subsection{Single function}

We assume we have a single function of the model output, denoted $F(Y)$.
The goal is to obtain sensitivities of $F$ with respect to the components
of $Z$.

Computation of the vector of sensitivities $ $
\begin{equation}
\tl F=\part FZ=\part F{X_{1}}\tl X_{1}+\dots+\part F{X_{N}}\tl X_{N}\label{eq:generalTL}
\end{equation}
 is done using the forward (tangent linear) mode. If we include the
trivial equations $\tl X_{1}=\tl X_{1},\cdots,\tl X_{N}=\tl X_{N}$,
then we can rewrite the combined expressions in matrix notation. This
will prove helpful when constructing the adjoint mode, using the fact
that adjoint of a matrix A is defined as its conjugate transpose (or
simply the transpose, if the matrix A has only real elements)
\begin{equation}
\left(\begin{array}{c}
\tl X_{1}\\
\dots\\
\tl X_{N}\\
\tl F
\end{array}\right)=\left(\begin{array}{ccc}
1 & \dots & 0\\
\dots & \dots & \dots\\
0 & \dots & 1\\
\part F{X_{1}} & \dots & \part F{X_{N}}
\end{array}\right)\left(\begin{array}{c}
\tl X_{1}\\
\dots\\
\tl X_{1}
\end{array}\right)\label{eq:generalTLmatrix}
\end{equation}

To compute each component of the vector $\part FZ$ we need to evaluate
the expression \eqref{eq:generalTL} a number of N times, every time
with a different input vector $\tl Z=\left\{ \tl X_{1},\dots,\tl X_{N}\right\} .$For
example, to compute the derivative with respect to $X_{j}$, the vector
$\tl Z$ has the value $\tl Z=\left(0,0,\dots,1,0,\dots,0\right)$,
with the only nonzero element in the $j-th$ position.

To construct the reverse (adjoint) mode, we start with the transposed
matrix equation \eqref{eq:generalTLmatrix} With the notation of $\adj A$
for the adjoint variable corresponding to an original variable $A$,
we have 
\[
\left(\begin{array}{c}
\adj X_{1}\\
\dots\\
\adj X_{N}
\end{array}\right)=\left(\begin{array}{cccc}
1 & \dots & 0 & \part F{X_{1}}\\
\dots & \dots & \dots & \dots\\
0 & \dots & 1 & \part F{X_{N}}
\end{array}\right)\left(\begin{array}{c}
\adj X_{1}\\
\dots\\
\adj X_{N}\\
\adj F
\end{array}\right)
\]
 Consequently, we obtain the following expressions
\begin{eqnarray}
\adj X_{1} & =\adj X_{1} & +\part F{X_{1}}\adj F\label{eq:generalAdj}\\
\dots & \dots & \dots\nonumber \\
\adj X_{N} & =\adj X_{N} & +\part F{X_{N}}\adj F\nonumber 
\end{eqnarray}

Thus we can obtain the sensitivities in one single run of \eqref{eq:generalAdj},
with $\adj F=1$ and the adjoint variables $\adj X_{1},\dots,\adj X_{N}$
initialized to zero.

\subsection{Composite functions}

We can generalize this procedure if the output is obtained through
evaluation of a composite function of P single functions (which is
in fact how the computer codes are represented): 
\[
F=F^{P}\circ F^{P-1}\circ\dots\circ F^{1}(Z)
\]
 We apply the tangent linear mode to each $F^{j}$, and we combine
them in the recursion relationship. For the adjoint (reverse mode),
we construct the adjoint for each $F^{j}$, and we combine them in
reverse order.

Let us describe the process using the matrix notation. If we view
the tangent linear as the result of the multiplication of a multiplication
of a number of operator matrices
\[
MAT=Mat_{1}\cdot Mat_{2}\cdots Mat_{P}
\]

where each matrix $Mat_{j}$ represents either a subroutine or a single
statement, then the adjoint approach can be viewed as a product of
adjoint subproblems
\[
Mat^{T}=Mat_{P}^{T}\cdot Mat_{P-1}^{T}\cdots Mat_{1}^{T}
\]

Let us describe how it works through an example for P=2. The computational
flow is described by the following diagram 
\[
Z\rightarrow F^{1}(Z)\rightarrow F^{2}\left(F^{1}(Z)\right)\rightarrow Y
\]

For simplicity, we denote by A the output of $F^{1}(Z)$ and by B
the output of $F^{2}\left(F^{1}\left(Z\right)\right)$ . $Y$ is the
scalar that is the final output. For example, Y can be the value of
the function to be calibrated (in the calibration setup) or, respectively,
the price of the option (in the pricing setup, e..g. using Monte Carlo
or finite differences). With these notations, we have
\[
Z\rightarrow A\rightarrow B\rightarrow Y
\]

Applying tangent linear methodology (essentially differentiation line
by line) we have
\begin{eqnarray*}
\tl A & = & \part AZ\tl Z\\
\tl B & = & \part BA\tl A\\
\tl Y & = & \part YB\tl B
\end{eqnarray*}

Putting everything together, we get
\[
\tl Y=\part YB\part BA\part AZ\tl Z
\]

Using notation from AD literature, the adjoint quantities $\adj Z,\adj A,\adj B,\adj Y$
denote the derivatives of $Y$ with respect to $Z,A,B$ and, respectively,
to $Y$ . We note that this implies that $\adj Y=1$. Differentiating
again, and with a superscript $^{T}$ denoting a matrix or vector
transpose, we obtain
\[
\adj Z=\left(\part YZ\right)^{T}=\left(\part YA\part AZ\right)^{T}=\left(\part AZ\right)^{T}\adj A
\]

In a similar way, we obtain 
\begin{eqnarray*}
\adj A=\left(\part YA\right)^{T}=\left(\part YB\part BA\right)^{T}=\left(\part BA\right)^{T}\adj B\\
\adj B=\left(\part YB\right)^{T}=\left(\part YB\right)^{T}\cdot1=\left(\part YB\right)^{T}\adj Y
\end{eqnarray*}

Putting everything together, we get
\[
\adj Z=\left(\part AZ\right)^{T}\left(\part BA\right)^{T}\left(\part YB\right)^{T}\adj Y
\]

We notice that the tangent linear approach proceeds forward (in forward
mode)through the process
\[
\tl Z\rightarrow\tl A\rightarrow\tl B\rightarrow\tl Y
\]

while the adjoint approach proceeds backwards (in reverse mode)
\[
\adj Y\rightarrow\adj B\rightarrow\adj A\rightarrow\adj Z
\]

\subsection{Checking the correctness of the implementation}

There are several checks that need to be made \cite{Navon_et_al_1992,Alexe_et_al_2009,Giles_et_al_2005}.

First, at any level of the code, the development of the discrete adjoint
model can be checked by appling the following identity
\[
\left(AQ\right)^{T}\left(AQ\right)=Q^{T}\left[A^{T}\left(AQ\right)\right]
\]

where $Q$ represents the input to original code, and $A$ represents
either a single statement or a subroutine

We compare the gradient computed using AD to the gradient computed
by Finite Difference (with a step size such that sufficient convergence
is obtained). We also compare the gradient computed using AD in Forward
mode versus the gradient computed using AD in Adjoint mode. We expect
those two gradients to be identical up to machine precision.

If possible, we may use complex numbers \cite{Giles_et_al_2005,Squire_Trapp_1998},
to avoid roundoff errors due to computations with finite difference.

\section{Simple example}

We want to compute the gradient of the function $f(a,b,c)=\left(w-w_{0}\right)^{2}$,
where $w$ is obtained using the following sequence of statements
\begin{eqnarray*}
\\
u & = & \sin\left(ab\right)+cb^{2}+a^{3}c^{2}\\
v & = & \exp\left(u^{2}-1\right)+a^{2}\\
w & = & ln\left(v^{2}+1\right)+cos\left(c^{2}-1\right)
\end{eqnarray*}

The input vector is denoted by by $z=\left\{ a,b,c\right\} $, intermediate
variables $u,w,v$ and output $f$. 

We show how the sensitivities with respect to a,b,c, namely 
\[
\frac{\partial f}{\partial a},\frac{\partial f}{\partial b},\frac{\partial f}{\partial c}
\]
 are computed in both forward and adjoint mode. For forward mode we
also write the expressions in matrix notation, to make it easier to
understand how the adjoint mode is constructed.

\subsection{Forward (tangent linear) mode}

We follow the computational flow, and thus we start by getting the
sensitivities with respect to the intermediate variables, denoted
by $\tl u,\tl v,\tl w$ 
\begin{eqnarray*}
\tl u & = & \part ua\tl a+\part ub\tl b+\part uc\tl c\\
\part ua & = & b\cos(ab)+3a^{2}c^{2}\\
\part ub & = & a\cos(ab)+2cb\\
\part uc & = & b^{2}+2a^{3}c
\end{eqnarray*}

Hence we obtain
\[
\tl u=\left[b\cos(ab)+3a^{2}c^{2}\right]\tl a+\left[a\cos(ab)+2cb\right]\tl b+\left[b^{2}+2a^{3}c\right]\tl c
\]

In matrix notation
\[
\left(\begin{array}{c}
\tl a\\
\tl b\\
\tl c\\
\tl u
\end{array}\right)=\left(\begin{array}{ccc}
1 & 0 & 0\\
0 & 1 & 0\\
0 & 0 & 1\\
b\cos(ab)+3a^{2}c^{2} & a\cos(ab)+2cb & b^{2}+2a^{3}c
\end{array}\right)\left(\begin{array}{c}
\tl a\\
\tl b\\
\tl c
\end{array}\right)
\]

In a similar way we obtain
\begin{eqnarray*}
\tl v & = & 2u\exp\left(u^{2}-1\right)\tl u+2a\tl a\\
\tl w & = & \frac{2v}{v^{2}+1}\tl v-2c\sin\left(c^{2}-1\right)\tl c\\
\tl f & = & 2\left(w-w_{0}\right)\tl w
\end{eqnarray*}

In matrix notation
\begin{eqnarray*}
\left(\tl v\right) & = & \left(\begin{array}{cc}
2a & 2u\exp\left(u^{2}-1\right)\end{array}\right)\left(\begin{array}{c}
\tl a\\
\tl u
\end{array}\right)\\
\left(\tl w\right) & = & \left(\begin{array}{cc}
-2c\sin\left(c^{2}-1\right) & \frac{2v}{v^{2}+1}\end{array}\right)\left(\begin{array}{c}
\tl c\\
\tl v
\end{array}\right)\\
\left(\tl f\right) & = & \left(2\left(w-w_{0}\right)\right)\left(\tl w\right)
\end{eqnarray*}

To obtain the required sensitivity with respect to $j-th$ component
of input vector$z$, we evaluate the above expressions starting with
$\tl z$ which has the $j-th$ component set to 1 and all the other
components set to 0.

More specifically, the sensitivities at a given set of variables $z^{(0)}=\left\{ a^{(0)},b^{(0)},c^{(0)}\right\} $
are computed by calling the forward mode with the initial value $\tl z$
defined as follows:
\begin{eqnarray*}
\part fa\left(z^{(0)}\right)=\tl f & computed\, with & \tl z=\left(\tl a,\tl b,\tl c\right)=\left(1,0,0\right)\\
\part fb\left(z^{(0)}\right)=\tl f & computed\, with & \tl z=\left(\tl a,\tl b,\tl c\right)=\left(0,1,0\right)\\
\part fc\left(z^{(0)}\right)=\tl f & computed\, with & \tl z=\left(\tl a,\tl b,\tl c\right)=\left(0,0,1\right)
\end{eqnarray*}

\subsection{Adjoint (reverse) mode}

To write the Adjoint, we need to take statements in reverse order.
Employing this approach to the statements of the forward mode yields
\begin{eqnarray*}
\left(\adj w\right) & = & \left(2\left(w-w_{0}\right)\right)\left(\adj f\right)\\
\left(\begin{array}{c}
\adj c\\
\adj v
\end{array}\right) & = & \left(\begin{array}{c}
-2c\sin(c^{2}-1)\\
\frac{2v}{v^{2}+1}
\end{array}\right)\left(\adj w\right)\\
\left(\begin{array}{c}
\adj a\\
\adj u
\end{array}\right) & = & \left(\begin{array}{c}
2a\\
2u\exp(u^{2}-1)
\end{array}\right)\left(\adj v\right)\\
\left(\begin{array}{c}
\adj a\\
\adj b\\
\adj c
\end{array}\right) & = & \left(\begin{array}{cccc}
1 & 0 & 0 & b\cos(ab)+3a^{2}c^{2}\\
0 & 1 & 0 & a\cos(ab)+2cb\\
0 & 0 & 1 & b^{2}+2a^{3}c
\end{array}\right)\left(\begin{array}{c}
\adj a\\
\adj b\\
\adj c\\
\adj u
\end{array}\right)
\end{eqnarray*}

Thus the adjoint (reverse) mode is constructed using the following
sequence of statements
\begin{eqnarray}
\adj w & = & 2\left(w-w_{0}\right)\adj f\label{eq:exampleAdjointMode}\\
\adj c & = & -2c\sin(c^{2}-1)\adj w\nonumber \\
\adj v & = & \frac{2v}{v^{2}+1}\adj w\nonumber \\
\adj a & = & 2a\adj v\nonumber \\
\adj u & = & 2u\exp(u^{2}-1)\adj v\nonumber \\
\adj a & = & \adj a+(b\cos(ab)+3a^{2}c^{2})\adj u\nonumber \\
\adj b & = & \adj b+(a\cos(ab)+2cb)\adj u\nonumber \\
\adj c & = & \adj c+(b^{2}+2a^{3})c\adj u\nonumber 
\end{eqnarray}

We can compute all 3 sensitivities of function $f$ with respect to
$a,b,c$ by a single application of the adjoint mode \eqref{eq:exampleAdjointMode},
with starting point $\adj z=1$. More specifically, we have
\begin{eqnarray*}
\part fa\left(z^{(0)}\right)=\adj a & computed\, with & \adj a=1\\
\part fb\left(z^{(0)}\right)=\adj b & computed\, with & \adj b=1\\
\part fc\left(z^{(0)}\right)=\adj c & computed\, with & \adj c=1
\end{eqnarray*}

The reader is encouraged to verify that the gradient computed via
the Forward mode or, respectively, the Reverse mode has identical
value.

\section{Example in PDE solver framework}

The following PDE is considered for $u(t,x)$, on the spatial domain$A\leq x\leq B$
\[
\begin{cases}
\part ut & =\part ux\\
u(0,x) & =u_{0}(x)\\
u(t,A) & =f(t)\\
u(t,B) & =g(t)
\end{cases}
\]

We discretize the PDE in space and time, for a spatial grid $\left\{ x_{j}\right\} $and
time grid $\left\{ T^{k}\right\} .$We use notation of superscript
for time index and subscript for spatial index. For simplicity of
exposition, we discretize using a central difference scheme in space,
a first order scheme in time, and we consider that both spatial grid
and, respective, temporal grid are constant, with $\Delta x$ and
$\Delta t$
\[
u_{j}^{k+1}=u_{j}^{k}+\frac{\Delta t}{2\Delta x}\left(u_{j+1}^{k}-2u_{j}^{k}+u_{j}^{k-1}\right)
\]

We denote by $c$ the ratio $\frac{\Delta t}{2\Delta x}$. With that,
and incorporating the boundary conditions, we have
\begin{eqnarray*}
u_{1}^{k+1} & = & f(T^{k+1})\triangleq f_{k+1}\\
u_{j}^{k+1} & = & u_{j}^{k}+c\left(u_{j+1}^{k}-2u_{j}^{k}+u_{j}^{k-1}\right)\,\,\,\,\,\, j=2...N\\
u_{N}^{k+1} & = & g(T^{k+1})\triangleq fg_{k+1}
\end{eqnarray*}

We want to minimize the difference $F\triangleq\sum_{j=1}^{N}\left(u_{j}^{M}-Y_{j}\right)^{2}$,
with $Y_{j}$ desired values to get at time T

We want to compute sensitivities of F with respect to the discretized
initial condition $\left\{ u_{j}^{0}\right\} ,$where $u_{j}^{0}\triangleq u_{0}(x_{j})$

\subsection{Forward (tangent linear) mode}

The tangent linear approach has the expression
\[
\tl u_{j}^{k+1}=\left(1-2c\right)\tl u_{j}^{k}+c\left(\tl u_{j+1}^{k}+\tl u_{j-1}^{k}\right)\,\,\,\,\,\,\, j=2...N
\]

In matrix notation
\begin{equation}
\left(\begin{array}{c}
\tl u_{1}^{k+1}\\
\dots\\
\tl u_{j-1}^{k+1}\\
\tl u_{j}^{k+1}\\
\tl u_{j+1}^{k+1}\\
\dots\\
\tl u_{N}^{k+1}
\end{array}\right)=\left(\begin{array}{ccccccc}
0 & \dots & \dots & \dots & \dots & \dots & 0\\
c & 1-2c & c & \dots & \dots & \dots & \dots\\
\dots & \dots & \dots & \dots & \dots & \dots & \dots\\
\dots & 0 & c & 1-2c & c & \dots & \dots\\
\dots & \dots & \dots & \dots & \dots & \dots & \dots\\
\dots & \dots & \dots & \dots & c & 1-2c & c\\
0 & \dots & \dots & \dots & \dots & \dots & 0
\end{array}\right)\left(\begin{array}{c}
\tl u_{1}^{k}\\
\dots\\
\tl u_{j-1}^{k}\\
\tl u_{j}^{k}\\
\tl u_{j+1}^{k}\\
\dots\\
\tl u_{N}^{k}
\end{array}\right)\label{eq:PDEmatrix}
\end{equation}

The last step is 
\[
\tl F=2\left(u_{1}^{M}-Y_{1}\right)\tl u_{1}^{M}+\dots+2\left(u_{j}^{M}-Y_{j}\right)\tl u_{j}^{M}+\dots+2\left(u_{N}^{M}-Y_{N}\right)\tl u_{N}^{M}
\]

In matrix notation
\begin{equation}
\left(\tl F\right)=\left(\begin{array}{ccccc}
2\left(u_{1}^{M}-Y_{1}\right) & \dots & 2\left(u_{j}^{M}-Y_{j}\right) & \dots & 2\left(u_{N}^{M}-Y_{N}\right)\end{array}\right)\left(\begin{array}{c}
\tl u_{1}^{M}\\
\dotsc\\
\tl u_{j}^{M}\\
\dots\\
\tl u_{N}^{M}
\end{array}\right)\label{eq:PDEF}
\end{equation}

\subsection{Adjoint (reverse) mode}

We go backwards. and the starting point is \eqref{eq:PDEF}. The corresponding
adjoint statements are
\begin{eqnarray*}
\adj u_{1}^{k} & = & 2\left(u_{1}^{M}-Y_{1}\right)\\
\dots & \dots & \dots\\
\adj u_{j}^{M} & = & 2\left(u_{j}^{M}-Y_{j}\right)\adj F\\
\dots & \dots & \dots\\
\adj u_{N}^{M} & = & 2\left(u_{N}^{M}-Y_{N}\right)\adj F
\end{eqnarray*}

Then we take the transpose of the matrix operator in \eqref{eq:PDEmatrix}.
The corresponding adjoint statements are, for $k=M,M-1,\dots,1,0$
\begin{eqnarray*}
\adj u_{1}^{k} & = & c\adj u_{2}^{k+1}\\
\adj u_{2}^{k} & = & (1-2c)\adj u_{2}^{k+1}+c\adj u_{3}^{k+1}\\
\dots & \dots & \dots\\
\adj u_{j}^{k} & = & c\adj u_{j-1}^{k+1}+(1-2c)\adj u_{j}^{k+1}+c\adj u_{j+1}^{k+1}\\
\dots & \dots & \dots\\
\adj u_{N-1}^{k} & = & c\adj u_{N-2}^{k+1}+(1-2c)\adj u_{N-1}^{k+1}\\
\adj u_{N}^{k} & = & c\adj u_{N}^{k+1}
\end{eqnarray*}

The required sensitivities are given by $\adj u_{j}^{0}$, for $j=1...N$

\section{Example in Monte Carlo framework (evolution of SDE)}

We consider the SDE for a N-dimensional X vector 
\[
dX=a(X,t)dt+\sigma(X,t)dW
\]

The initial value for X, at time t=0, is denoted by$X^{0}=\left(X_{1}(T^{0}\right),...,X_{N}(T^{0})$)

For simplicity of exposition, we consider that we discretize it using
Euler-Maruyama scheme, with time points $T^{k},k=1...M$ and $T^{0}=0$
\begin{eqnarray*}
X(T^{k+1})-X(T^{k}) & = & a(X(T^{k}),T^{k})\Delta t+\sigma(X(T^{k}),T^{k})\Delta W^{k}\\
\Delta T^{k} & = & T^{k+1}-T^{k}\\
\Delta W^{k} & = & \varepsilon\cdot\sqrt{T^{k+1}-T^{k}}
\end{eqnarray*}

where the random number is chosen from $\mathcal{N\mbox{(0,1)}}$

We can recast this discretization scheme into the following expression
\begin{equation}
X(T^{k+1})=\Phi(X(T^{k}),\Theta)\label{eq:SDEfunc}
\end{equation}

where $\Phi$ may also depend on a set of model parameters $\Theta=\left\{ \theta_{1},...,\theta_{P}\right\} $

We also consider that we have to price an option with payoff $G(X(T))$

We want to compute price sensitivities ({}``Greeks'') with respect
to $X^{0}$ and, respectively, to the model parameters $\Theta$

\subsection{Forward (tangent linear) mode}

We consider first the sensitivities with respect to initial conditions
\begin{equation}
\part{G(X(T))}{X_{i}(T^{0})}=\sum_{j=1}^{N}\part{G(X(T))}{X_{j}(T)}\part{X_{j}(T)}{X_{i}(T^{0})}=\sum_{j=1}^{N}\part{G(X(T))}{X_{j}(T)}\Delta_{ij}(T^{0},T)\label{eq:SDEIC}
\end{equation}

where we have used notation $\Delta_{ij}(T^{0},T^{k})\triangleq\part{X_{j}(T^{k})}{X_{i}(T^{0})}$

We rewrite \eqref{eq:SDEIC} in vector matrix notation
\[
\left[\part{G(X(T))}{X(T^{0})}\right]^{\mathrm{T}}=\left[\part{G(X(T))}{X(T)}\right]^{\mathrm{T}}\cdot\Delta(T^{0},T)
\]

For simplicity we write the previous relationship in the following
format, using the fact that $T^{M}=T$ 
\[
\left[\part GX[0]\right]^{\mathrm{T}}=\left[\part GX[M]\right]^{\mathrm{T}}\cdot\Delta[M]
\]

where the superscript $^{T}$ denotes the transpose

Differentiating \eqref{eq:SDEfunc} yields
\begin{eqnarray}
\part{X\left(T^{k+1}\right)}{X\left(T^{k}\right)}=\part{\Phi\left(X(T^{k}),\Theta\right)}{X\left(T^{k}\right)}\label{eq:SDEone}\\
\Rightarrow\part{X\left(T^{k+1}\right)}{X\left(T^{0}\right)}=\part{X\left(T^{k+1}\right)}{X\left(T^{k}\right)}\part{X\left(T^{k}\right)}{X\left(T^{0}\right)} & =\part{\Phi\left(X(T^{k}),\Theta\right)}{X\left(T^{k}\right)}\part{X\left(T^{k}\right)}{X\left(T^{0}\right)} & =D[k]\part{X\left(T^{k}\right)}{X\left(T^{0}\right)}\nonumber 
\end{eqnarray}

where $D[k]\triangleq\part{\Phi\left(X(T^{k}),\Theta\right)}{X\left(T^{k}\right)}$

We rewrite \eqref{eq:SDEone} as 
\[
\Delta[k+1]=D[k]\cdot\Delta[k]
\]

Then we have an iterative process
\begin{eqnarray}
\left[\part GX[0]\right]^{\mathrm{T}} & = & \left[\part GX[M]\right]^{\mathrm{T}}\cdot\Delta[M]=\left[\part GX[M]\right]^{\mathrm{T}}\cdot D[M-1]\cdot\Delta[M-1]\label{eq:SDETgOne}\\
 & = & \left[\part GX[M]\right]^{\mathrm{T}}\cdot D[M-1]\cdot D[M-2]\cdot\Delta[M-2]=\dots\nonumber \\
 & = & \left[\part GX[M]\right]^{\mathrm{T}}\cdot D[M-1]\cdot\cdots D[1]\cdot\Delta[1]\nonumber \\
 & = & \left[\part GX[M]\right]^{\mathrm{T}}\cdot D[M-1]\cdot\cdots D[0]\cdot\Delta[0]\nonumber 
\end{eqnarray}

The matrix $\Delta[0]$ is the identity matrix, since
\begin{equation}
\Delta[0]=\left(\part{X_{j}(T^{0})}{X_{k}(T^{0})}\right)_{jk}=\left(\begin{array}{cccc}
1 & \cdots & \cdots & 0\\
\cdots & \cdots & \cdots & \cdots\\
\cdots & \cdots & 1 & 0\\
0 & \cdots & 0 & 1
\end{array}\right)\label{eq:deltaZeroIdentity}
\end{equation}

The iterations in \eqref{eq:SDETgOne} employ matrix-matrix product.
We will see later that the adjoint mode will involve matrix-vector
product instead, which will offer important computational savings.

Now we move to the sensitivities with respect to model parameters
\begin{equation}
\part{G(X(T))}{\theta_{p}}=\sum_{j=1}^{N}\part{G(X(T))}{X_{j}(T)}\part{X_{j}(T)}{\theta_{p}}=\sum_{j=1}^{N}\part{G(X(T))}{X_{j}(T)}\Psi_{jp}(T)\label{eq:SDETgParam}
\end{equation}

where we have used notation $\Psi_{jp}(T^{k})\triangleq\part{X_{j}(T^{k})}{\theta_{p}}$

Differentiating \eqref{eq:SDEfunc} with respect to parameters yields
\begin{equation}
\part{X\left(T^{k+1}\right)}{\Theta}=\part{\Phi\left(X(T^{k}),\Theta\right)}{X\left(T^{k}\right)}\part{X(T^{k})}{\Theta}+\part{\Phi\left(X(T^{k}),\Theta\right)}{\Theta}\label{eq:SDEOneParam}
\end{equation}

Making the notations $D[k]\triangleq\part{\Phi\left(X(T^{k}),\Theta\right)}{X\left(T^{k}\right)}$
and $B[k]\triangleq\part{\Phi\left(X(T^{k}),\Theta\right)}{\Theta}$
for the corresponding matrices, we rewrite \eqref{eq:SDEOneParam}
as 
\[
\Psi[k+1]=D[k]\cdot\Psi[k]+B[k]
\]

Then we have an iterative process
\begin{eqnarray}
\left[\part{G(X(T))}{\Theta}\right]^{\mathrm{T}}=\left[\part{G(X(T))}{X(T)}\right]^{\mathrm{T}}\cdot\Psi[M]=\left[\part{G(X(T))}{X(T)}\right]^{\mathrm{T}}\cdot\left(D[M-1]\cdot\Psi[M-1]+B[M-1]\right)\label{eq:TgParamFinal}\\
=\left[\part{G(X(T))}{X(T)}\right]^{\mathrm{T}}\cdot\left(D[M-1]\cdot\left(D[M-1]\cdot\Psi[M-1]+B[M-1]\right)+B[M-1]\right)=\cdots=\nonumber \\
=\left[\part{G(X(T))}{X(T)}\right]^{\mathrm{T}}\cdot(B[M-1]+D[M-1]\cdot B[N-2]+\cdots+D[M-1]\cdots D[1]\cdot B[0])\nonumber 
\end{eqnarray}

\subsection{Adjoint (reverse) mode}

We construct first the adjoint for computing the sensitivities with
respect to initial condition. We start with the adjoint equation
\begin{equation}
V[k]=D^{T}[k]\cdot V[k+1]\label{eq:AdjStart}
\end{equation}

where the superscript $^{T}$ denotes the transpose

In a recursive manner we obtain
\begin{equation}
V[0]=D^{T}[0]\cdot V[1]=D^{T}[0]\cdot D^{T}[1]\cdot V[2]=\cdots=D^{T}[0]\cdot D^{T}[1]\cdot\cdots D^{T}[M-1]\cdot V[M]\label{eq:SDEAdjOne}
\end{equation}

By taking transpose of \eqref{eq:SDEAdjOne} we have
\begin{equation}
V^{T}[0]=V^{T}[M]\cdot D[M-1]\cdot D[M-2]\cdot\cdots D[1]\cdot D[0]\label{eq:SDEAdjTwo}
\end{equation}

We set $V[M]$ to the value $\left(\part GX[M]\right)^{T}$and we
combine \eqref{eq:SDEAdjTwo} and \eqref{eq:SDETgOne}, which gives
\[
\part GX[0]=V^{T}[0]\cdot\Delta[0]
\]

But the matrix $\Delta[0]$ is the identity matrix, according to \ref{eq:deltaZeroIdentity}

Thus we obtain the sensitivities with respect to initial conditions,
namely $\part GX[0]$ by applying the recursive relationship \eqref{eq:SDEAdjTwo}
to find $V^{T}[0]$. We note that the product in this iterative process
is of the matrix-vector type, not matrix-matrix as it was for tangent
linear mode 

Now we move to sensitivities with respect to model parameters

We use again the adjoint equation \eqref{eq:AdjStart}. With same
initial condition for $V[M]$ , namely $V[M]=\left(\part GX[M]\right)^{T}$,
and evolving from time $T^{M-k}$ to time $T^{M}$ we have that
\[
\part{G(X(T))}{\Theta}\cdot D[M-1]\cdot D[M-2]\cdot\cdots D[M-k]\cdot B[M-k-1]=V[M-k]
\]
 Thus we can rewrite \eqref{eq:TgParamFinal} as
\begin{equation}
\part{G(X(T))}{\Theta}=\sum_{k=0}^{M-1}V^{T}[k+1]\cdot B[k]\label{eq:SDEAdjParamFinal}
\end{equation}

The values of adjoint vectors $V[k]$ were computed as part of the
adjoint approach for sensitivities with respect to initial conditions.

The values of $B[k]$ can be precomputed. We may be able to compute
them analytically (e.g., if the parameters are volatilities and vol
surface is assumed to have a certain parametric representation, such
as cubic spline), otherwise we can employ the adjoint procedure.

\section{Example in Monte Carlo framework (copula)}

We follow a procedure similar to the one described in \cite{Capriotti_Giles_2010}

Let us describe the setup. We have a state vector $\mathbb{\mathbb{X}}$
of N components, an instrument with a payoff $P(\mathbb{X})$, and
the probability distribution $\mathbb{Q\mbox{\ensuremath{\left(\mathbb{\mathbb{X}}\right)}}}$
according to which the components of $\mathbb{\mathbb{X}}$ are distributed.
For simplicity we consider a N-dimensional Gaussian copula to model
the co-dependence between the components of the state vector, namely
a joint cumulative density function of the form
\[
\Phi_{N}\left[\Phi_{1}\left(\varphi_{1}\left(X_{1}\right),\cdots,\varphi_{N}\left(X_{N}\right)\right);\rho\right]
\]

where $\Phi_{N}\left[Z_{1},\cdots,Z_{N};\rho\right]$ is a N-dimensional
multivariate Gaussian distribution with zero mean and a correlation
matrix $\rho$, $\Phi^{-1}$ is the inverse of the standard normal
cumulative distribution and $\varphi_{i}\left(X_{i}\right)$, i=1...N
are the marginal distributions of the underlying components.

The option value is obtained through averaging of Monte Carlo sampling
\[
V=\frac{1}{N_{MC}}\sum_{j=1}^{N_{MC}}P\left(\mathbb{X}{}^{(i)}\right)
\]

where the superscript $^{(i)}$ denotes the $i-$th Monte Carlo path.

The sampling of the N jointly distributed normal random variables
$\left(Z_{1},Z_{2},\cdots,Z_{N}\right)$ can be done using several
approaches, such as Cholesky factorization, spectral or singular value
decomposition. We select Cholesky factorization

$\rho=C\cdot C\mathrm{^{T}}$ because it will enable us to use an
existing procedure for its adjoint. Starting from a vector $\hat{Z}$
of independent standard normal variables, we obtain a vector $Z$of
jointly normal random variables distributed according to $\Phi_{N}\left[Z_{1},\cdots,Z_{N};\rho\right]$
through the product
\[
Z=C\cdot\hat{Z}
\]

We also use the following:
\begin{itemize}
\item if $Z_{i}$ is sampled from a standard normal distribution then $\Phi\left(Z_{i}\right)$
is in $\mathcal{{U\left[\mbox{0,1}\right]}}$
\item if $X_{i}$ is distributed according to marginal distribution $\varphi_{i}$,
then $\varphi\left(X_{i}\right)$ is in $\mathcal{{U\left[\mbox{0,1}\right]}}$
\end{itemize}
Then $X_{i}=\varphi^{-1}\left(\Phi\left(Z_{i}\right)\right)$ is distributed
according to marginal distribution $\varphi_{i}$

Therefore we have the following algorithm \cite{Capriotti_Giles_2010}
\begin{enumerate}
\item Generate a vector $\Xi$ of independent standard normal variables
\item Correlate the components through $Z=C\cdot\Xi$
\item Set $U_{i}=\Phi\left(Z_{i}\right),\,\, i=1...N$ 
\item Set $X_{i}=\varphi_{i}^{-1}\left(\Phi\left(Z_{i}\right)\right)=\varphi^{-1}\left(U_{i}\right),\,\, i=1...N$
\item Calculate the payoff estimator $P\left(X_{1},...,X_{N}\right)$
\end{enumerate}
We now show how sensitivities can be computed in this setup. 

The correlation matrix is an input to the procedure, so we may compute
correlation risk, i.e., sensitivities of the price with respect to
entries in the correlation matrix

These marginal distributions may be obtained from a set of $ $$N^{MARG}$
discrete call, put, digital call and digital put values (which may
be given as corresponding implied volatilities).We may also compute
sensitivities with respect to those inputs, denoted by $\varpi_{j},j=1...N^{MARG}$

\subsection{Forward (tangent linear) mode}

Assuming that the payoff function is regular enough (e.g., Lipschitz
continuous) the standard pathwise differentiation corresponds to forward
(tangent linear) mode. The differentiation is applied to steps 1-5
in the above algorithm. We need to pay attention if any given step
is dependent (implicitly or explicitly) on the input parameters with
respect to which we want to compute the sensitivities. Step 1 stays
the same. Step 2 and 3 are differentiated if we want correlation risk,
otherwise they remain unchanged. Steps 4 and 5 are differentiated
regardless which of the two types of sensitivities of sensitivities
we want to compute.

To differentiate Step 4 we start from $X_{i}=\varphi_{i}^{-1}\left(U_{i}\right)\Rightarrow U_{i}=\varphi_{i}(X_{i})$. 

Differentiating the last equality gives the following formula, if
we have the propagated derivative $\tl U_{i}$ of the variable $U_{i}$
(i.e., we compute the correlation risk) 
\[
\tl U_{i}=\part{\varphi_{i}}x\left(X_{i}\right)\tl X_{i}\Rightarrow\tl X_{i}=\frac{\tl U_{i}}{\part{\varphi_{i}}x\left(X_{i}\right)}
\]

If we need to compute sensitivities with respect to $\varpi_{j}$,
then differentiating the same equality gives
\[
\tl X_{i}=\frac{\tl{\varpi}_{j}}{\part{\varphi_{i}}x\left(X_{i}\right)}
\]

We now present the algorithm for tangent linear mode. For ease of
presentation we write explicitly the algorithm only for the case of
correlation sensitivities. 

We assume that we want to compute the sensitivity with respect to
entry $\rho_{lk}$ of the correlation matrix
\begin{enumerate}
\item Generate a vector $\Xi$ of independent standard normal variables
\item Calculate $\tl Z=\tl{C_{lk}}\cdot\tl{\Xi}$, where $\tl{C_{lk}}$
is the sensitivity of Cholesky factor $C$ with respect to $\rho_{lk}$
\item Set $\tl U_{i}=\part{\Phi}x\left(Z_{i}\right),\,\, i=1...N$
\item Set $\tl X_{i}=\frac{\tl U_{i}}{\part{\varphi_{i}}x\left(X_{i}\right)},\,\, i=1...N$ 
\item Calculate $\tl P=\sum_{i=1}^{N}\part P{X_{i}}\left(X_{i}\right)\cdot\tl X_{i}$
\end{enumerate}
We note that the derivative of the marginal distribution, denoted
by $\part{\varphi_{i}}x\left(X_{i}\right)$, is the probability density
function associated with the marginal distribution $\varphi_{i}$,
while the derivative $\part{\Phi}x$ is the standard normal probability
density function

\subsection{Adjoint (reverse) mode}

The adjoint mode consists of the adjoint counterparts for each of
the steps in the forward mode, plus the adjoint of Cholesky factorization
\cite{Smith_1995}. The computation is done in reverse order

The resulting algorithm consists of the 5 steps described above (in
the forward mode) plus the following steps corresponding to adjoint
counterparts
\begin{enumerate}
\item Calculate the adjoint of the payoff estimator $\adj X_{k}=\part P{X_{k}}(X_{k}),\:\, k=1...N$
\item Calculate $\adj U_{k}=\frac{\adj X_{k}}{\part{\varphi_{i}}x\left(\varphi_{k}^{-1}\left(U_{k}\right)\right)},\,\, k=1...N$
\item Calculate $\adj Z_{k}=\adj U_{k}\part{\Phi}x\left(Z_{k}\right),\,\, k=1...N$
\item Calculate $\adj C=\adj Z\cdot\Xi^{\mathrm{T}}$
\end{enumerate}
The matrix $\adj C=\left(\adj C_{ij}\right)_{i,j=1..N}$ obtained
at the end of the procedure contains all derivatives of payoff estimator
with respect to entries $\varrho_{ij}$ of the correlation matrix.
We can see that all these sensitivities were computed by running the
adjoint mode only once, as opposed to the forward (tangent linear)
mode, which had to be run separately for each entry in the correlation
matrix (with a total number of runs of $N^{2}$)

\section{Example in calibration/optimization framework}

Let us consider that we have the same SDE as in the previous chapter.
We want to minimize a cost functional of the type 
\[
F=\sum\left(ModelPrice[j]-MarketPrice[j]\right)^{2}
\]

with variables to be optimized being given by model parameters $\Theta=\left\{ \theta_{1},...,\theta_{P}\right\} $

The gradient of the cost functional with respect to the model parameters
would have the expression
\[
\left(\begin{array}{c}
\cdots\\
\part F{\theta_{k}}\\
\cdots
\end{array}\right)=\left(\begin{array}{c}
\cdots\\
2\sum\left(ModelPrice[j]-MarketPrice[j]\right)\part{\left(ModelPrice[j]\right)}{\theta_{k}}\\
\cdots
\end{array}\right)
\]

The adjoint procedure enables us to compute $ $$\part{\left(ModelPrice[j]\right)}{\theta_{k}}$
in a similar way to \eqref{eq:SDEAdjParamFinal}

We note here that the above procedure assumes implicitly the existence
of the gradient of the functional. It is our experience \cite{Homescu_Navon_2003_b}that
the discrete adjoint can still be applied for cases such gradient
doe not exist; in those cases the numerical adjoint code will provide
us not the gradient, but rather subgradients. Consequently, one will
have to employ optimization algorithms that are especially designed
to use subgradients instead of gradient

\section{Computational finance literature on adjoint and AD}

In the last several years quite a few papers were added to the literature
on adjoint/AD applied to computational finance \cite{Beveridge_et_al_2010,Capriotti_et_al_2011,Capriotti_Giles_2010,Capriotti_Giles_2011,Chan_Joshi_2010,Chan_Joshi_2010b,Denson_Joshi_2009,Denson_Joshi_2010,Denson_Joshi_2010b,Giles_Glasserman_2006,Joshi_Chan_2009,Joshi_Pitt_2010,Joshi_Wiguna_2011,Joshi_Yang_2010,Joshi_Yang_2010b,Joshi_Yang_2011,Kaebe_2010,Spilda_2010,Achdou_Pironneau_2005,Bastani_Guerreri_2008,Capriotti_2011,Stehle_2010}
. For selected papers we give an overview in the following sections

\subsection{Computation of Greeks}

A major part of the literature related to this topic is due to Giles,
Capriotti and, respectively, Joshi (and their collaborators). 

The seminal paper of \cite{Giles_Glasserman_2006} applied the adjoint
approach to the computation of Greeks (Delta and Vega) for swaptions
using pathwise differentiation method in the context of LIBOR Market
Model (LMM). The portfolio considered had portfolio had 15 swaptions
all expiring at the same time, N periods in the future, involving
payments/rates over an additional 40 periods in the future. Interested
in computing Deltas, sensitivity to initial N+40 forward rates, and
Vegas, sensitivity to initial N+40 volatilities. The computational
efficiency was improved by a factor of 10 when the forward method
was employed instead of finite differences. Then the adjoint approach
reduced the cost by several orders of magnitude, compared to the forward
approach: for N=80 periods, by a factor of 5 for computation of Deltas
only, and respectively by a factor of 25 for computation of both Deltas
and Vegas. 

This was extended in \cite{Leclerc_et_al_2009} to the pricing of
Bermudan-style derivatives. For testing they have used fi{}ve 1xM
receiver Bermudan swaptions (M = 4, 8, 12, 16, 20) with half-year
constant tenor distances. The speedup was as follows (for M=20): by
a factor of 4 for only Deltas, and by factor of 10 for both Deltas
and Vegas.

The pathwise approach is not applicable when the financial payoff
function is not differentiable. To address these limitations, a combination
the adjoint pathwise approach for the stochastic path evolution with
likelihood ratio method (LRM) for the payoff evaluation is presented
in \cite{Giles_2007,Giles_2008}. This combination is termed {}``Vibrato''
Monte Carlo. The Oxford English Dictionary describes \textquotedblleft{}vibrato\textquotedblright{}
as \textquotedblleft{}a rapid slight variation in pitch in singing
or playing some musical instruments\textquotedblright{}. The analogy
to Monte Carlo methods is the following; whereas a path simulation
in a standard Monte Carlo calculation produces a precise value for
the output values from the underlying stochastic process, in the vibrato
Monte Carlo approach the output values have a narrow probability distribution.

Applying concepts of adjoint/AD for correlation Greeks were considered
in \cite{Capriotti_Giles_2010}. The pricing of an instrument based
on N underlyings is done with Monte Carlo within a Gaussian copula
framework, which connects the marginal distributions of the underlying
factors. The sampling of the N jointly normal random variables is
efficiently implemented by means of a Cholesky factorization of the
correlation matrix. Correlation risk is obtained in a highly efficient
way by implementing the pathwise differentiation approach in conjunction
with AD, using the adjoint of Cholesky factorization. Numerical tests
on basket default options shows a speedup of 1-2 orders of magnitude
(100 times faster than bumping for 20 names, and 500 times for 40
names).

Examples of pathwise differentiation combined with AD are shown in
\cite{Capriotti_2011}. For a basket option priced in a multidimensional
lognormal model, the savings are already larger than one order of
magnitude for medium sized baskets (N=10). For {}``Best of'' Asian
options in a local volatility setting, the savings are reported to
over one order of magnitude for a relatively small number (N=12) of
underlying assets.

Adjoint algorithmic differentiation can be used to implement efficiently
the calculation of counterparty credit risk \cite{Capriotti_et_al_2011}.
Numerical results show a reduction by more than two orders of magnitude
in the computational cost of Credit Valuation Adjustment (CVA). The
example considered is a portfolio of 5 swaps referencing distinct
commodities Futures with monthly expiries with a fairly typical trade
horizon of 5 years, the CVA bears non-trivial risk to over 600 parameters:
300 Futures prices, and at the money volatilities, (say) 10 points
on the zero rate curve, and 10 points on the Credit Default Spread(CDS)
curve of the counterparty used to calibrate the transition probabilities
of the rating transition model. required for the calculation of the
CVA. The computational time for CVA sensitivities is less than 4 times
the computational time time spent for the computation of the CVA alone,
as predicted by AD theory. As a result, even for this very simple
application, adjoint/AD produces risk over 150 times faster than finite
differences: for a CVA evaluation taking 10 seconds, adjoint approach
produces the full set of sensitivities in less than 40 seconds, while
finite differences require approximately 1 hour and 40 minutes.

In the framework of co-terminal swap-rate market model \cite{Joshi_Yang_2009}
presented an efficient algorithm to implement the adjoint method that
computes sensitivities of an interest rate derivative (IRD) with respect
to different underlying rates, yielding a speedup of a factor of 15
for Deltas of a Bermudan callable swaption with rates driven by a
5-factor Brownian motion. They show a total computational order of
the adjoint approach of order $O(nF)$, with $n$ the number of co-terminal
swap rates, assumed to be driven by number$F$ Brownian motions, compared
to the order of standard pathwise method which is $O(n^{3})$. It
was extended to Vegas in \cite{Joshi_Yang_2011}, in the context of
three displaced diffusions swap-rate market models. A slight modification
of the method of computing Deltas in generic market models will compute
market Vegas with order $O(nF)$ per step. Numerical results for the
considered tests show that computational cots for Deltas and Vegas
will be not more than 1.5 times the computational cots of only Deltas.
CMS spread options in the displaced-diffusion co-initial swap market
model are priced and hedged in \cite{Joshi_Yang_2010a}. The evolution
of the swap-rates is based on an efficient two-factor predictor-corrector(PC)
scheme under an annuity measure. Pricing options using the new method
takes extremely low computational times compared with traditional
Monte Carlo simulations. The additional time for computing model Greeks
using the adjoint method can be measured in milliseconds. Mostly importantly,
the enormous reduction in computational times does not come at a cost
of poorer accuracy. In fact, the prices and Greeks produced by the
new method are sufficiently close to those produced by a one-step
two-factor PC scheme with 65,535 paths.

In the framework of the displaced-diffusion LIBOR market model, \cite{Joshi_Wiguna_2011}
compared the discretization bias obtained when computing Greeks with
pathwise adjoint method for the iterative predictor-corrector and
Glasserman-Zhao drift approximations in the spot measure to those
obtained under the log-Euler and predictor-corrector approximations
by performing tests with interest rate caplets and cancellable receiver
swaps. They have found the iterative predictor-corrector method to
be more accurate and slightly faster than the predictor-corrector
method, the Glasserman-Zhao method to be relatively fast but highly
inconsistent, and the log-Euler method to be reasonably accurate but
only at low volatilities.

In the framework of the cross-currency displaced-diffusion LIBOR market
model, \cite{Beveridge_et_al_2010} employs adjoint techniques to
compute Greeks for two commonly traded cross-currency exotic interest
rate derivatives: cross-currency swaps (CCS) and power reverse dual
currency (PRDC) swaps. They measure the computational times relative
to the basic implementation of the crosscurrency LIBOR market model,
that is with standard least squares regression used to compute the
exercise strategy. It was reported that, using the adjoint pathwise
method, the computational time for all the Deltas and Vega for each
step and the exchange Vega is only slightly larger compared to the
computational time required to compute one Delta using finite differences.

Adjoint approach was applied for computation of higher order Greeks
(such as Gammas) in \cite{Joshi_Yang_2010b} and \cite{Joshi_Yang_2010},
where they show that the Gamma matrix (i.e. the Hessian) can be computed
in AM + B times the number of operations where M is the maximum number
of state variables required to compute the function, and A,B are constants
that only depend on the set of floating point operations allowed..
In the first paper numerical results demonstrate that the computing
all $\nicefrac{n(n+1)}{2}$ Gammas for Bermudan cancellable swaps
in the LMM takes roughly $\nicefrac{n}{3}$ times as long as computing
the price. In the second paper numerical results are shown for an
equity tranche of a synthetic CDO with 125 names. To compute all the
125 finite-difference Deltas, the computational time for doing so
will be 125 times of the original pricing time (if we use forward
difference estimates). However, it only takes up to 1.49 times of
the original pricing time if we use algorithmic differentiation. There
are 7,875 Gammas to estimate. If one uses central difference estimates,
it will take 31,250 times of the original pricing time to achieve
this. It only takes up to 195 times to estimate all the Gammas with
algorithmic Hessian methods.

In the framework of Markov-functional models, \cite{Denson_Joshi_2010}
demonstrates how the adjoint PDE method can be used to compute Greeks.
This paper belongs to the ContAdj category. It obtains an adjoint
PDE, which is solved once to obtain all of the model Greeks. In the
particular case of a separable LIBOR Markov-functional model the price
and 20 Deltas, 40 mapping Vegas and 20 skew sensitivities (80 Greeks
in total) of a 10-year product are computed in approximately the same
time it takes to compute a single Greek using finite difference. The
instruments considered in the paper were: interest rate cap, cancellable
inverse floater swap and callable Bermudan swaption.

In the framework of Heston model, the algorithmic differentiation
approach was considered in \cite{Chan_Joshi_2010} to compute the
first and the second order price sensitivities.. Issues related to
the applicability of the pathwise method are discussed in this paper
as most existing numerical schemes are not Lipschitz in model inputs.
While AD/adjoint approach is usually considered for primarily providing
a very substantial reduction in computational time of the Greeks,
this paper also shows how its other major characteristic, namely accuracy,
can be extremely relevant in some cases. Computing price sensitivities
is done using the Lognormal (LN) scheme, the Quadratic- Exponential
(QE) scheme, the Double Gamma (DG) scheme and the Integrated Double
gamma (IDG) scheme. Numerical tests show that the sample means of
price sensitivities obtained using the Lognormal scheme and the Quadratic-Exponential
scheme can be highly skewed and have fat-tailed distribution while
price sensitivities obtained using the Integrated Double Gamma scheme
and the Double Gamma scheme remain stable.

The adjoint/AD approach is also mentioned in the very recent {}``opus
magna'' on interest rate derivatives \cite{Andersen_Piterbarg_2010_vI,Andersen_Piterbarg_2010_vII,Andersen_Piterbarg_2010_vIII},
written by 2 well-known practitioners.

\subsection{Calibration}

To the best of our knowledge, \cite{Achdou_Pironneau_2005} was the
first time that an adjoint approach was presented for calibration
in the framework of computational finance

In \cite{Kaebe_et_al_2009} adjoint methods are employed to speed
up the Monte Carlo-based calibration of financial market models. They
derive the associated adjoint equation /(and thus are in ContAdj category)
and propose its application in combination with a multi-layer method.
They calibrate two models: the Heston model is used to confirm the
theoretical viability of the proposed approach, while for a lognormal
variance model the adjoint-based Monte Carlo algorithm reduces computation
time from more than three hours (for finite difference approximation
of the gradient)to less than ten minutes

The adjoint approach is employed in \cite{Maruhn_2010,Maruhn_2011}
to construct a volatility surface and for calibration of local stochastic
volatility models. The fitting of market bid/ask-quotes is not a trivial
task, since the quotes differ in quality as well as quantity over
timeslices and strikes, and are furthermore coupled by hundreds of
arbitrage constraints. Nevertheless, the adjoint approach enabled
on the fly fitting (less than 1 second) for real-life situations:
a volatility matrix with 17 strikes and 8 maturities. 

AD is applied to calibration of Heston model in\cite{Hakala_2011},
through differentiation of vanilla pricing using characteristic functions.
It takes advantage of the fact that it can be applied to complex numbers
(if they are {}``equipped'' with auto-differentiation). Applying
AD to numerical integration algorithm is straightforward, and the
resulting algorithm is fast and accurate.

In \cite{Turinici_2009} the adjoint approach (of ContAdj type) is
applied to calibrate a local volatility model

Calibration of a local volatility model is also shown in \cite{Spilda_2010}

\section{AD and adjoint applied within a generic framework in practitioner
quant libraries}

As more financial companies become familiarized with the advantages
offered adjoint/AD approach when applied to computational finance,
this approach will be more and more integrated in their quant libraries.
Very recent presentations \cite{Capriotti_2011a,Capriotti_2010,Hakala_2011,Maruhn_2011,Baxter_2010}
at major practitioner quant conferences indicate that this effort
is under way at several banks, such as Credit Suisse, Unicredit, Nomura,
etc. Within this framework one may imagine a situation where Greeks
are computed (using AD techniques) in real time to provide hedging
with respect to various risk factors: interest rate, counterparty
credit, correlation, model parameters etc.

Due to complexity of quant libraries, AD tools and special techniques
of memory management, checkpointing etc can prove extremely useful,
potentially leveraging knowledge acquired during AD applications to
large software packages in other areas: meteorology, computational
fluid dynamics etc where such techniques were employed \cite{Griewank_Walther_2008,Charpentier_Ghemires_2000,Hascoet_Dauvergne_2008,Alexe_et_al_2009,Giles_et_al_2005,Bischof_et_al_2005,Bartlett_et_al_2006}

\subsection{Block architecture for Greeks}

\cite{Baxter_2010} presents a {}``Block architecture for Greeks''
, using calculators and allocators. Sensitivities are computed for
each block component, including root-finding and optimization routines,
trees and latices.

Here is an example included in the presentation. First some notations:
every individual function (routine), say $f:\mathbb{R^{\mbox{m}}\rightarrow\mathbb{R^{\mbox{n}}\mbox{\,\,\,\ y=f(x)}}}$is
termed a {}``calculator'' function, and a corresponding {}``allocator''
function is defined as 
\[
G_{f}:\mathbb{R^{\mbox{n}}\rightarrow\mathbb{R^{\mbox{m}}\mbox{\,\,\,\,}G_{\mbox{f}}\left(\mbox{\ensuremath{\part Vy}}\right)=}\mbox{\ensuremath{\part Vy}}\mbox{\ensuremath{\cdot}J\ensuremath{_{f}}}}
\]

If we have a sequence of functions f, g, h, then we calculate the
Greeks by starting with the end point $\part VV=1$ and working backwards
by calling the allocator functions one-by-one in reverse order to
the original order of the function calls. 

Suppose we have a pricing routine with three subroutines
\begin{itemize}
\item f : Interpolates market data onto the relevant grid from input market
objects
\item g : performs some calibration step on the gridded market data
\item h : core pricer which uses the calibrated parameters
\end{itemize}
We need to write the three risk allocator functions for these three
subroutines. Then we feed the starting point $\part VV=1$ into the
pricer\textquoteright{}s allocator to get back the vector of risks
to the calibrated parameters. The next step is to feed this into the
calibrator\textquoteright{}s allocator to get back the vector of risks
to the gridded market data. Finally, we feed that risk vector into
the interpolator\textquoteright{}s allocator to get back risk to the
original input market objects.

\subsection{Real Time Counterparty Risk Management in Monte Carlo}

\cite{Capriotti_2011a} presents a framework for Real Time Counterparty
Risk Management, employing AD to compute Credit Valuation Adjustment
(CVA), with a reduction in computational time from 100 min (using
bump and reprice) to 10 seconds, for a a portfolio of 5 commodity
swaps over a 5 years horizon (over 600 risks)

\section{Additional resources}

Additional resources include reference books \cite{Bischof_et_al_2008,Bucker_et_al_2006,Corliss_et_al_2002,Griewank_Walther_2008},
results on matrix differentiation \cite{Giles_2008a,Petersen_Pedersen_2008,Smith_1995},
the online community portal for Automatic Differentiation \cite{Autodiff},
various AD tools such as FASTOPT, FASTBAD, ADMAT, FAD, ADOL-C, RAPSODIA,
TAPENADE \cite{ADTools}, papers on rules to construct the adjoint
code \cite{Griewank_Walther_2008,Bischof_et_al_2003,Giering_Kaminski_1998,Giering_Kaminski_2002,Cusdin_Muller_2003,Cusdin_Muller_2004,Bischof_1995,Achdou_Pironneau_2005,Smith_2000}

\section{Conclusion}

We have presented an overview of adjoint and automatic(algorithmic)
differentiation techniques in the framework of computational finance,
and illustrated the major advantages of this approach for computation
of sensitivities: great reduction of computational time, improved
accuracy, and a systematic process to construct the corresponding
{}``adjoint'' code from existing codes

\bibliographystyle{plain}
\bibliography{References_Adjoints_Arxiv}

\end{document}